\documentstyle[prl,aps,epsf,multicol]{revtex}
\begin{document}

%%%%%%%%%%%FIGURES%%%%%%%%%%%%%%%%%%
%
\newcommand{\fig}[2]{\epsfxsize=#1\epsfbox{#2}}
% 
%
%%%%%%%%%DEUX COLONNES%%%%%%%%%%%%%%
%   
\newcommand{\passage}{%%
\end{multicols}\widetext\noindent\rule{8.8cm}{.1mm}%
  \rule{.1mm}{.4cm}} 
 \newcommand{\retour}{%%
 %        \hspace{.2cm}
\noindent\rule{9.1cm}{0mm}\rule{.1mm}{.4cm}\rule[.4cm]{8.8cm}{.1mm}%
         \begin{multicols}{2} }
 \newcommand{\unecol}{\end{multicols}}
 \newcommand{\deuxcol}{\begin{multicols}{2}}
%
%
%%%%%%%%%%%%%%%%%%%%%%%%%%%%%%%%%%%%
%
%
\newcommand{\beq}{\begin{equation}}
\newcommand{\eeq}{\end{equation}}
\newcommand{\beqa}{\begin{eqnarray}}
\newcommand{\eeqa}{\end{eqnarray}}

\tolerance 2000

\author{Daniel S. Fisher{$^1$}, Pierre Le Doussal{$^2$} and Cecile 
Monthus{$^3$} 
}
\address{{$^1$} Lyman Laboratory of Physics, Harvard University, 
Cambridge MA 
02138, USA}
\address{{$^2$}CNRS-Laboratoire de Physique Th\'eorique de 
l'Ecole Normale Sup\'erieure \cite{email2},
24 rue Lhomond,75231 Cedex 05, Paris France. }
\address{{$^3$} Service de Physique Th\'eorique, CEA Saclay, 
F91191 Gif sur Yvette France}

\title{Random Walks, Reaction-Diffusion, and Nonequilibrium Dynamics of Spin 
Chains in One-dimensional Random 
Environments}
% \date{\today}
\maketitle

\begin{abstract}
Sinai's model of diffusion in one-dimension with random local bias is studied by 
a real space renormalization group which yields asymptotically exact long time 
results.
The distribution of the position of a particle and
the probability of it not returning to the origin are obtained, as well as 
the two-time distribution which exhibits "aging" with $\frac{\ln t}{\ln t'}$  
scaling
and a singularity at $\ln t =\ln t'$. The effects of a small uniform force are 
also studied. Extension to motion of  
many domain walls yields non-equilibrium time dependent correlations for the 1D 
random field Ising model with
Glauber dynamics and "persistence" exponents of 1D 
reaction-diffusion models with random forces. 

\end{abstract}

%\pacs{to be added}

%\narrowtext
\deuxcol

The development of order in systems with a broken symmetry is of 
interest in 
many contexts. "Coarsening" of domain structures evolving towards
equilibrium has been studied extensively \cite{bray}. But little
is known analytically about domain growth in the presence of quenched 
disorder
  \cite{anderson,fisher_huse}.
Nevertheless, phenomenological descriptions of the non-equilibrium 
dynamics of 
various random magnetic systems have 
been developed in terms of "droplets" separated by domain walls 
\cite{fisher_huse}. Due to the very slow dynamics associated with 
activation 
over large free energy barriers, even the apparent equilibrium properties of 
these systems are dominated by the non-equilibrium dynamics, as also 
occurs in 
infinite-range models  \cite{Cuku,fisher_huse}.
Even in one-dimension some random systems exhibit  ultra slow growth 
and 
aging phenomena. Exact results in 1D could thus be used as testing 
grounds
for more complex $D>1$ cases which have resisted analytic attack.

In this Letter we study 
the diffusion of a particle in
a 1D random potential which itself has the statistics of a 1D random walk 
\cite{sinai}.  Extensions to many interacting particles allows us to study, 
via 
domain walls,
the Glauber dynamics of 
1D Ising models, in particular random field ferromagnets and spin glasses 
in a 
magnetic field. This leads also to the consideration of more general
diffusion-reaction processes in such
energy landscapes. 
Various analytic results are known for the single particle model (Sinai 
model)
\cite{sinai,biased,kesten,derrida_pomeau,ledou_1d,monthus_diffusion,laloux_pld_sinai}
but these primarily concern single time quantities.
Here we use a real space renormalization group
(RSRG) method related to that used to study
disordered quantum spin chains \cite{ma_dasgupta,danfisher_rg2,derrida_phi4}.
This allows us to compute a
host of quantities such as first passage (persistence) exponents,
single time correlations and even two time correlations that are probed in 
aging 
experiments.  
Despite its approximate character, the RSRG
yields many results that should be exact. 

The model is defined as follows: Particles diffuse on a 1D lattice in a 
potential 
 ${U_i}$, with $i$ the  site index.
A ``force'' variable $f_i \equiv  U_i - U_{i+1}$
is defined on each bond $(i,i+1)$ 
with these $f_i$ independent
random variables. Since one can  group together neighboring bonds with 
the
same sign of the force, we study
with no loss of generality,  a "zigzag" potential (see Fig. 
\ref{fig1})
with the $f_i$ alternatively positive
and negative but with a distribution of "bond" lengths $l_i$.
Our  model is thus defined by 
$f_i= (-1)^{i+1} F_i$ where the positive $F_i=|U_i - U_{i+1}|$, which are 
effectively energy barriers, are
the natural variables. 
The pairs of bond variables $F,l$ are chosen independently
from bond to bond from a distribution $P(F,l)$.
In the presence of a directionality bias one needs two distinct
distributions
$P(F,l)$ for ``descending bonds''
and $R(F,l)$ for ``ascending bonds'', both 
normalized to unity.

\begin{figure}[htb]
\centerline{ \fig{6cm}{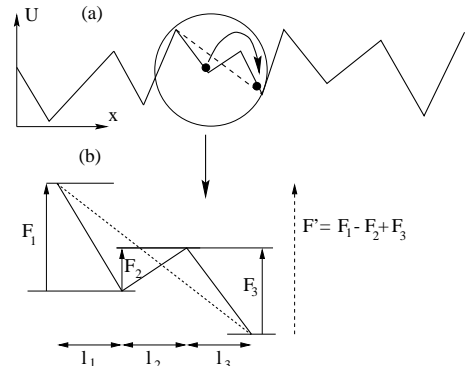} }
\caption{
{\narrowtext (a) Energy landscape in Sinai model (b) 
decimation method: the bond with the smallest barrier
$F_{min}=F_2$ is eliminated resulting in three bonds being grouped into
one.\label{fig1} }}
\end{figure}
We are interested in long times when the behavior will be dominated by 
large barriers
and it is on these that we must focus.  Our RG procedure is conceptually  
simple:
in a given energy
landscape it consists of  iterative decimation of 
the bond with the {\it smallest barrier}, say $F_2=U_3 - U_2$
as illustrated in Fig. \ref{fig1}. At time scales much longer than $\exp 
(F_2/T)$, 
local equilibrium will be established between sites 2 and 3 and the rate 
for the 
walker to get from 4 to 1 will be essentially the same as it would be if 
sites 2 
and 3 did not exist but 1 and 4 were instead connected by a bond with 
barrier 
$F'=F_1 - F_2 + F_3$ and length $l'=l_1 + l_2 + l_3$. We thus carry out 
exactly 
this replacement which preserves the zigzag structure and the  larger 
scale 
extrema
of the potential. 
With $\Gamma$ defined to be the smallest remaining barrier at a given 
stage of 
the RG, we eliminate the barriers in the range
$\Gamma<F<\Gamma +d\Gamma$. The new variables are {\it independent} 
from bond to 
bond.
 Introducing the variable
$\zeta\equiv F-\Gamma$ one finds the following RG flow equations for 
the
probabilities $P(\zeta,l,\Gamma)$ and $R(\zeta,l,\Gamma)$ (with 
$0<\zeta<\infty$):
\begin{eqnarray}   \label{rg1}
(\partial_\Gamma - \partial_\zeta) P &=&
R(0,.)*_l P *_{\zeta l} P + (P^\Gamma_0 - R^\Gamma_0) P
\nonumber \\
(\partial_\Gamma - \partial_\zeta) R &=&
P(0,.)*_l R *_{\zeta l} R + (R^\Gamma_0 - P^\Gamma_0) R
\end{eqnarray}
Here $*_\zeta$ denotes a convolution with respect
to $\zeta$ only and $*_{\zeta,l}$ with respect to both $\zeta$ and $l$ and we
define
 $P^\Gamma_0 \equiv \int_0^{\infty} dl P(\zeta=0,l,\Gamma)$
and similarly for $R^\Gamma_0$.
The dynamics implied by this RG is rather simple.  Making the obvious 
identification of $\Gamma = T\ln (t/t_0)$ from Arrhenius dynamics,
we see that at very long time scales the 
renormalized landscape consists entirely of deep valleys separated by 
high 
barriers. A good approximation to the long time dynamics is thus to place 
the walker {\it at the bottom} of the renormalized valley at scale $T\ln t$ in 
which 
it was initially, since, with high probability, it will be near to that point 
\cite{sinai}. 
  Upon proper rescaling of space and time this 
approximation 
becomes
exact as $\Gamma$ tends to $\infty$ as was proven in ref. \cite{sinai} 
for the unbiased case. 
It remains valid in the weakly biased case in the limit 
that 
the 
bias parameter that controls the long time properties, $\mu$, defined 
implicitly for the original model with unit length bonds by $<\exp (-\mu 
f_i/T)> 
= 1$, is very small (see \cite{derrida_pomeau,ledou_1d}). 

The RG equations (\ref{rg1}) are identical to those derived 
for the random transverse field Ising chain (RTFIC)  in \cite{danfisher_rg2}  
with the identification of $\ln h_k = F_{2 k}$ and 
$\ln J_k = F_{2k+1}$ in the 
RTFIC with the ascending and descending barriers
respectively \cite{random_dirac}. Thus duality in the RTFIC corresponds
to reversing the average force. Criticality then corresponds to
the zero drift case, while the Griffiths phase in the RTFIC \cite{danfisher_rg2}
corresponds to the biased phase with zero velocity 
\cite{derrida_pomeau,ledou_1d}. The deviation from criticality parameter
\cite{danfisher_rg2}
$\delta \equiv  
(\langle \ln h\rangle - \langle \ln J\rangle) / [var(\ln h) + var(ln J)]$
is analogous at small $\delta$ to
$\mu /2$.

We consider first the long time dynamics of a single particle, 
starting with the symmetric, zero bias, case that has the same 
distribution for all the bonds; i.e., $R = P$.  
For large $\Gamma$ , the distribution flows to one of a 
family of scaling solutions of the 
RG equations (\ref{rg1}). The rescaled probability
$\tilde{P}(\eta,\lambda) = (\Gamma^3/\sigma) P(\eta \Gamma, \lambda 
\Gamma^2/\sigma, \Gamma)$
in terms of the rescaled variables $\eta\equiv \zeta/\Gamma$,
$\lambda\equiv l\sigma /\Gamma^2$, when Laplace transformed in 
$\lambda$ to $s$, 
is found to be\cite{danfisher_rg2} 
$\tilde{P}(\eta,s) = (\sqrt{s}/\sinh (\sqrt{s}))
\exp(-\eta\sqrt{s}\coth(\sqrt{s}))$. The average bond length is 
$\overline{l} = 
\frac{1}{2\sigma}\Gamma^2$ and we recover the scaling 
$x \sim \ln^2 t$ \cite{sinai}. 
 The large scale variance of the potential $<(U_i-U_j)^2> 
\approx 
2\sigma |l_{i-j}|$, with $l_{i-j}$ the distance from i to j is 
conserved 
by the RG, fixing $\sigma$. 

The fact that the renormalized barrier distribution becomes infinitely broad 
is 
the source of the exactness of our long time results.  At late stages of the 
RG, 
the chances that two neighboring bonds have $F$'s that are within order 
$T$ of 
each other tends to zero for large $\Gamma$.  Thus substantial errors 
that are 
introduced by assigning a particle to one of two almost-equal-depth 
neighboring 
valleys rather than splitting its distribution between the two valleys will 
occur rarely
at long scales.  Furthermore, any such error is wiped out by a 
later decimation 
which eliminates the two valleys in favor of a deeper valley.  Since 
in a deep renormalized valley, the particle tends to be very close to 
the bottom on the scale of $\overline{l}(\Gamma)$ \cite{sinai,kesten}, we can obtain the 
scaled 
distribution of 
the position of a particle at time $t$ that started at the origin at time zero, 
directly from $P(l,\Gamma = T\ln t)$. We henceforth set 
$T=1$ 
and measure distances in units such that $\sigma=1$. The  dynamics "within" one 
bond and errors made in 
early stages of the RG will generally only change the microscopic cutoff time  
in $\ln t$'s.

The renormalized dynamics corresponds to moving 
the particle from its starting point (distributed uniformly on a  bond) to the 
lower-potential end of the bond.  
The distribution of its position at time $t$ 
averaged over the ensemble of random potentials is thus 
$\overline{\mbox{Prob}(x,t|0,0)}= 
\int_{|x|}^{\infty} dl P(l,\Gamma)/\overline{l}(\Gamma)$.
With $\Gamma=\ln t$, it takes the scaling form 
$\overline{\mbox{Prob}(x,t|0,0)}=\frac{1}{\ln^2 t} q(\frac{x}{\ln^2 t})$
where
\begin{eqnarray} \label{kesten}
q(X) = \frac{4}{\pi} 
\sum_{n = 0}^{\infty} \frac{(-1)^{n}}{2 n+ 1}
e^{- \frac{1}{4} \pi^2 |X| (2 n + 1)^2}
\end{eqnarray}
With $\sigma$ reinserted, this 
coincides with Kesten's rigorous result \cite{kesten} for a Brownian potential, 
as it should \cite{sinai}.

But we can now generalize to the biased case with a small average 
potential 
drop per unit length $2\delta >0$.  The RG flows Eq(\ref{rg1}) now involve 
the 
two  
distributions $R$ and $P$. The asymptotic behavior of these flows   
was found in  \cite{danfisher_rg2}; it obtains in the scaling limit that 
$\Gamma$ 
is large, while $\lambda = l/\Gamma ^2$ and $\gamma \equiv  \Gamma 
\delta$ are 
both fixed but arbitrary. In terms of the Laplace transform from $|X|\equiv 
|x|/\Gamma ^2$ to $s$, we obtain for the generalization of Eq(\ref{kesten}),
\begin{eqnarray} \label{biased}
&& q(X,\gamma) = (\frac{\gamma}{\sinh \gamma} )^2 [\theta (X) LT^{-1} 
\frac{1}{s}(1- \frac{\kappa 
e^{-\gamma }}{\kappa \cosh \kappa - \gamma\sinh \kappa}) \nonumber 
\\ 
&& + \theta (-X) LT^{-1} \frac{1}{s}
(1- \frac{\kappa e^\gamma }{\kappa \cosh \kappa + \gamma\sinh \kappa}) 
],
\end{eqnarray}
with $\kappa \equiv \sqrt{s+\gamma ^2}$ and the two terms arising 
from descending and ascending renormalized bonds, respectively.
In the limit of small $\gamma$, the behavior reduces to the symmetric 
case 
Eq(\ref{kesten}).  But for large $\gamma$, i.e $\ln t >> 1/\delta$, 
the distribution is heavily concentrated to the right of the origin being 
simply
$\overline{\mbox{Prob}(x,t|0,0)} \approx \theta (x) \exp (-x/\overline{x(t)}) / 
\overline{x(t)}$ with the mean displacement $\overline{x(t)} \approx 
t^{2\delta 
} / 
(4\delta ^2)$, consistent with the small $\mu$ limit
of the known \cite{biased,ledou_1d} "Levy front" $L_{\mu}(t/(\mu^2 x)^{1/\mu})$, 
although the exponent $\mu$ of the  anomalous diffusion, $x\sim t^\mu$ is 
correct only to leading order in $\delta$,
due to corrections to scaling neglected in our 
RG. We find that the model renormalizes onto a directed 
model
with traps (ascending bonds) of "release time" distribution $\rho(\tau) 
\sim
\tau^{-(1+\mu)}$ \cite{ledou_1d}.
 
Our method also enables us to compute two time quantities,
e.g. $B( x,t,x',t')\equiv \overline{\mbox{Prob}(xt|x't'|00)}$
which contains information about the dynamics after the system has 
"aged"
 from $t=0$ to $t'$,  but the full calculation
and result \cite{us_long} are too complicated to reproduce here. In the 
regime
$\Gamma=\ln t$ and $\Gamma'=\ln t'$ large with $\alpha  \equiv \ln t/\ln 
t'$
a fixed number, we obtain a scaling form 
$B\approx B_\alpha (\tilde{x},\tilde{x}')$ in the 
rescaled variables $\tilde{x}=x/\Gamma^2$ and
$\tilde{x}'=x'/\Gamma^2$. Our two time correlations thus
exhibit
a $\ln t'/\ln t$ aging regime, as found numerically in \cite{laloux_pld_sinai} 
and
argued in \cite{fisher_huse} for 
spin glasses in higher dimensions. Interestingly,
the rescaled distribution has a delta function 
component at the origin, suggested in \cite{laloux_pld_sinai}
but obtained here analytically. It 
arises from 
bonds not decimated between $t'$ and $t$,
i.e., from particles staying within the same
valley. The probability $D_{\Gamma, 
\Gamma'}(\zeta)$
that a particle on a bond with $F$ at $\Gamma$ has not moved from 
$\Gamma'$ to 
$\Gamma$
satisfies a linear RG equation which, when integrated, yields the
probability that a walker has not 
moved substantially between $t'$ and $t$
\begin{eqnarray}
D(t,t') = (\frac{\ln t'}{\ln t})^2
(\frac{5}{3} - \frac{2}{3} e^{-(\frac{\ln t}{\ln t'}-1)})
\end{eqnarray}
The mean square additional displacement  $\langle |x(t)-x(t')|^2 
\rangle$ at large $\alpha$ is $\approx \frac{61}{180}  (\ln t)^4$. 
But when $\ln t$ and $\ln t'$ are not too separated, i.e $\alpha \approx 1$
it is only $\approx (\ln t')^4
\frac{272}{315} (\ln t/\ln t' -1)$. Typically in this regime
the particle is trapped in a deep well, but  
there is a probability of order 
$(\Gamma -\Gamma ')/\Gamma '$ that one of the barriers of the well at 
time 
$t'$ is less than $\Gamma $.  If this does occur, then the walker will 
jump to the bottom of a deeper valley a distance of order 
$\overline{l}(\Gamma ') \sim \Gamma '^2$ away. Note, however, that for 
$(t-t')/t'=O(1)$ or less, this will not occur as $t'\to\infty$ and the behavior 
will instead be dominated by rare 
configurations -absent in the scaling limit- in which the valley at time t 
has 
two almost degenerate minima. Jumping between such minima  
persists even for $t \to \infty$ 
with 
$t-t'$ fixed and in this limit the statistics of the infinitely deep valley 
potential becomes that of a random walk restricted to have
$U_i-U_{valley-min}>0$ \cite{sinai,us_long,laloux_pld_sinai}.

We now turn to problems involving many walkers.
The Glauber dynamics of the 1D (classical) random field Ising model
corresponds to {\it two types} of domain walls
A and B which see {\it opposite} random  potentials with the forces $f_i$ 
being 
simply twice the corresponding random fields on the dual lattice sites. 
When the 
random fields are much smaller than the exchange $J$, the long time 
behavior for 
$T<<J$ will be universal. We focus on the evolution  starting from random 
initial conditions - e.g., after a quench from a high temperature.
At low temperatures, an A  wall quickly falls to the bottom of a valley 
only to
move to the bottom of a neighboring lower 
valley 
when $\ln t$ reaches the barrier height of the intervening bond. Likewise, 
B 
walls move from top
to top of "mountains". When an A and a B meet, they 
annihilate preserving the alternating ABAB
sequence. Analytic treatment is difficult because
decimation  generates correlations among the A and B positions.
Performing the RSRG numerically on a large sample \cite{us_long} 
we find that the system evolves to a state 
with one A at each minimum and one B at each maximum of the 
renormalized 
landscape. 
We thus make the Ansatz that this is the correct form of the asymptotic 
states. 
The RG analysis is then again simple \cite{derrida_phi4}.  The equal time spin 
correlations 
can be 
obtained from the difference between the probabilities that an even or 
an odd number of extrema of the renormalized potential -i.e., domain 
walls - 
exist between a given pair of points. 
For a symmetric distribution of random fields, we obtain:
\begin{eqnarray} \label{Ising}
\overline{\langle S_0(t)S_{L}(t) \rangle}\approx \sum_{n=-\infty }^{\infty } 
\frac{48 + 32 (2n+1)^2 \pi^2 X}{(2n+1)^4\pi^4} 
e^{-(2n+1)^2 \pi^2 X} \nonumber
\end{eqnarray}
with $X=\frac{L}{\Gamma^2}$, distances normalized as earlier and $\Gamma = T\ln 
t$.
At sufficiently long times $\Gamma >\Gamma_J  = 2J$, we can no 
longer ignore creation of pairs of walls. But, at this point, the energy 
cannot 
be lowered further by {\it any} process.  Thus if the renormalization is 
stopped at $\Gamma_J$, in 
the small field, low $T$ scaling limit the configuration of the walls 
corresponds, up 
to negligible thermal fluctuations, to the equilibrium state. The above
equation should then give the mean equilibrium spin correlation 
function with lengths measured in units of the Imry-Ma length above 
which the 
random fields dominate the exchange. 

Since 1D Ising spin glasses in a field are equivalent via a gauge 
transformation 
to 
random field ferromagnets,
we can also obtain results for such a system. If a large magnetic 
field is quickly reduced to be $<<J$ but non-zero, the domain wall 
dynamics will 
be like that for the ferromagnet with domain walls initially at 
every 
extremum of the potential.  The decay of the magnetization for 
log-times up to $\Gamma_J$ is given by the difference between the 
probabilities 
that a spin has flipped an even or an odd number of times.  We obtain 
$\overline{\langle S_i(t) \rangle} \sim \overline{l}(t)^{-\lambda}$ with 
$\lambda 
= 
\frac{1}{2}$.
Note that this value of $\lambda$ saturates the lower bound of 
$d/2$ in contrast to the pure 1D Ising case which saturates the upper 
bound of 
$\lambda =d$ \cite{fisher_huse}. For the symmetric RFIM one similarly 
finds 
that $\overline{\langle S_i(t) S_i(t') \rangle} 
\sim
(\overline{l}(t')/\overline{l}(t))^{\frac{1}{2}}$.

We next study "persistence" properties. One 
must now carefully distinguish between the {\it effective 
dynamics} (i.e  the walker jumping between valley bottoms )
and the {\it real} dynamics. The probability $N(t)$
that a {\it single walker} has {\it never} crossed its starting point 
$x(0)=x_0$ between $0$ and $t$ can be found by placing an absorbing boundary at 
$x_0$ and using methods similar to the calculation of
the endpoint magnetization in the RTFIC \cite{danfisher_rg2}.
We find $N(t) \sim \overline{l}(t)^{-\theta_1}$ at large times
with $\theta_1=\frac{1}{2}$ (c.f. $\theta_1=1$ in the pure case). A related 
quantity, $M(t)$, is the fraction of starting points, $x_0$, for which the {\it 
thermally averaged
position} $\langle x(t) | x(0)=x_0 \rangle$ never crosses $x_0$
up to time $t$. While in a single "run" in a given environment the walker 
typically crosses its starting point many times while trapped in a 
valley,  averaging over many runs in the same environment yields a $\langle x(t) 
\rangle$ which crosses $x_0$ exactly once each time the bond on which $x_0$ lies 
is decimated since this causes its valley bottom to 
cross $x_0$. At long times, the probability $M(t)$ thus reflects the effective 
dynamics, in particular the probability that the bond on which $x_0$ lies has 
{\it 
never} been decimated before time $t$ yielding $M(t) \sim 
\overline{l}(t)^{-\overline{\theta}_1}$
with
$\overline{\theta}_1=\frac{3-\sqrt{5}}{4}$.  Indeed, in the biased
case, the probability of no return of $\langle x(t) \rangle$ is like the 
spontaneous 
magnetization in the RTFIC, i.e. 
$M(t) \sim |\delta|^\beta$ for small $\delta$ with $\beta=\frac{3-\sqrt{5}}{2}$   
\cite{danfisher_rg2}. 

 More generally, in the effective dynamics, the probability of exactly $n$ 
crossings of the origin  up to time $t$ scales as $\ln(\ln t)$ in the unbiased 
case. The 
rescaled variable
$g=n/\ln (\ln t)$ has a multifractal distribution
$\mbox{Prob}(g) \sim  \overline{l}(t)^{-\overline{\theta}_1(g)}$ with
\begin{eqnarray}
 \label{multi} 
\overline{\theta}_1(g) = \frac{g}{2} \ln[2 g(g+\sqrt{g^2+\frac{5}{4}} )] + 
\frac{3}{4} - \frac{g}{2}
- \frac{1}{2} \sqrt{g^2+\frac{5}{4}}.
\end{eqnarray}
Since $\overline{\theta}_1(\frac{1}{3})=0$,
$g=\frac{1}{3}$ with probability $1$ at large times.
For a given walker, $\Xi(t) \equiv \frac{1}{t} \int_0^t 
x(\tau) d\tau$ will typically behave like $\langle x(t) \rangle$.  We 
conjecture that
the probability of $n=g \ln(\ln t)$ sign changes of $\Xi$ up to time $t$ decays 
with the same exponent $\overline{\theta}_1(g)$ for $g \leq \frac{1}{3}$. For 
larger $g$, the behavior is dominated by rare valleys with almost degenerate 
minima on opposite sides of the origin which yield extra sign changes in 
$\Xi(t)$. 

The persistence properties of the RFIM can similarly be analyzed. The 
probability $\Pi(t) \sim
\overline{l}(t)^{-\theta}$ that a given {\it spin} at $x=0$
has never flipped up to time $t$ is equal to the probability
that neither the nearest domain wall on one side, nor the nearest (opposite 
type) 
domain wall on the other side have crossed $x=0$. Assuming the nature of the 
asymptotic state is as  described earlier, we find
$\theta= 2 \theta_1 =1$ (c.f. $\theta=\frac{3}{4}$ in the pure case
\cite{derrida_hakim}).
In contrast,  the decay of the probability
that an initial {\it domain} survives up to time $t$, is $S(t) \sim 
\overline{l}(t)^{-\psi}$ with
 $\psi= \frac{3-\sqrt{5}}{4}=0.191$ (c.f. $\psi=0.252$ in the pure case 
\cite{krapivsky_psi}). 

Finally, one can study a broad class of reaction diffusion models
where all particles diffuse in the {\it same} unbiased energy landscape
and react or annihilate upon meeting, for example, 
identical particles $A$ which react as
$ A + A \to A$  with probability $1-r$ or annihilate
$ A + A \to \emptyset$  with probability $ r$. The fraction of the valleys with 
no particle 
in them tends to $p_{\emptyset}$, the stationary probability for the reaction 
process upon merging two valleys. Generalizing the absorbing boundary
method we obtain that, very generally, the probability
that $x=0$ has not been crossed by {\it any} particle
up to $t$ decays with the exponent
$\theta=1-p_{\emptyset}$, in our example, $\theta(r)= 1/(1+r)$.
It is interesting to note that the corresponding
exponent $\overline{\theta}(r)$ associated with
the thermally averaged particle trajectories is the 
solution of the hypergeometric equation
\cite{us_long}:
\begin{eqnarray}
\overline{\theta} U(-r/(1+r), 2 \overline{\theta},1) = U(-r/(1+r), 2 
\overline{\theta} + 1,1).
\end{eqnarray}
Remarkably, this $\overline{\theta}(r)$ 
is very close numerically to half the exact pure system result
\cite{derrida_hakim}
$\frac{1}{2} \theta_{pure}(r)
=-\frac{1}{8} + \frac{2}{\pi^2} (\arccos(\frac{r-1}{\sqrt{2} 
(r+1)}))^2$.

To conclude, we have applied a RSRG method to $1D$ random walks in the
presence of static random forces and obtained exact results for
coarsening dynamics, diffusion reaction models, and aging phenomena;
surprisingly by simpler means than for the corresponding pure models.
Extensions and details of the  present results will be given in
\cite{us_long}. 

This work has been supported in part (DSF) by the NSF via grants DMR 9630064, 
DMS 9304580 and Harvard University's MRSEC.

\vspace{20cm}.

\unecol

%\end{references}


\begin{thebibliography}{999}

%\begin{references}





\bibitem{email2}  Laboratoire associ\'e \`a l'ENS
et \`a l'Universit\'e Paris-Sud.

\bibitem{bray} A. J. Bray Adv. Phys. {\bf 43} 357 (1994).

\bibitem{anderson} see e.g. S.R. Anderson Phys. Rev B {\bf 36} 8435 
(1987).

\bibitem{fisher_huse} D.S. Fisher, D.A. Huse Phys. Rev. B {\bf 38} 373 
(1988).

\bibitem{Cuku} L. F. Cugliandolo and J. Kurchan; Phys. Rev. Lett. {\bf 
71}, 173
(1993) and J. Phys. {\bf A27}, 5749 (1994).

\bibitem{fisher_huse} D.S. Fisher, D.A. Huse Phys. Rev. B {\bf 38} 373 
(1988).

\bibitem{sinai} Y. G. Sinai Theor. Probab. Its Appl. {\bf 27} 247 (1982)
A.O. Golosov, Commun. Math.Phys 92, 491 (1984).

\bibitem{biased}
H. Kesten, M. Koslov, F. Spitzer, Compos. Math. {\bf30} 145 (1975).
H. Tanaka, Int. Conf. Math. (1994).

\bibitem{kesten} H. Kesten, Physica {\bf 138} A 299 (1986).

\bibitem{derrida_pomeau} B. Derrida, J. Stat. Phys. {\bf 31} 433 (1983).

\bibitem{ledou_1d} J. P. Bouchaud, A. Comtet, A. Georges,
P. Le Doussal Europhys. Lett. {\bf 3} 653 (1987),
Ann Phys. {\bf 201} 285 (1990).

\bibitem{monthus_diffusion} A. Comtet, J. Desbois and C. Monthus
Ann. Phys. {\bf 239} 312 (1995).  C. Monthus et al. Phys. Rev. E {\bf 54}
231 (1996).

\bibitem{laloux_pld_sinai}
L. Laloux and P. Le Doussal cond-mat/9705249 Phys. Rev. E 1997 (in 
press) and references therein.

\bibitem{ma_dasgupta} C. Dasgupta and 
S.K. Ma Phys. Rev. B {\bf 22} 1305 (1980); D.S. Fisher, Phys. Rev. B {\bf 50} 
3799 (1994).

\bibitem{danfisher_rg2}
D.S. Fisher, Phys. Rev. B {\bf 51} 6411 (1995).

\bibitem{derrida_phi4} Coarsening of the pure $T=0$ 1D $\phi^4$ model is 
somewhat similar as the smallest
domain is iteratively decimated; see
A.J. Bray, B. Derrida Phys. Rev. E {\bf 51}
R1633 (1995) and references therein.

\bibitem{us_long} D.S. Fisher, P. Le Doussal, C. Monthus
in preparation.

\bibitem{fokker_dirac}
A. Comtet, A. Georges, P. Le Doussal, Physics Lett. B {\bf 208} 487 (1988)

\bibitem{random_dirac} The RSRG yields correspondences
for the low energy (large time) properties. More formal
derivations can be made using the equivalence of random dirac operators 
and 1D diffusion
(see e.g L. Balents and M. P. A. Fisher
cond-mat/9706069  ; E. Witten Nucl. Phys. B {\bf 188} 513 (1981)
and Ref. \cite{fokker_dirac,ledou_1d}.

\bibitem{derrida_hakim} B. Derrida, V. Hakim, V. Pasquier
Phys. Rev. Lett. {\bf 75} 751 (1995) computed $\theta$
for the pure q-states Potts model, equivalent to the $r=1/(q-1)$
reaction diffusion process.

\bibitem{krapivsky_psi} P. L. Krapivsky, E. Ben Naim
Phys. Rev. E {\bf 56}
3788 (1997).

\end{thebibliography}
\end{document}